# Ferroelectric Domains and Evolution Dynamics in Twisted CuInP$_2$S$_6$ Bilayers

*Dongyu Bai[1†], Junxian Liu[1†], Yihan Nie[2], Yuantong Gu[1], Dongchen Qi[3], Arkady Krasheninnikov[4], Liangzhi Kou[1]\**

1. School of Mechanical, Medical and Process Engineering, Queensland University of Technology, Brisbane, Queensland 4001, Australia

2. College of Civil Engineering and Architecture, Zhejiang University, Hangzhou 310058, China

3. School of Chemistry & Physics, Queensland University of Technology, Brisbane, Queensland 4001, Australia

4. Institute of Ion Beam Physics and Materials Research, Helmholtz-Zentrum Dresden-Rossendorf, 01328 Dresden, Germany

Corresponding author email: *Liangzhi.kou@qut.edu.au*

**Abstract**

Polar domains and their manipulation—particularly the creation and dynamic control—have garnered significant attention, owing to their rich physics and promising applications in digital memory devices. In this work, using density functional theory (DFT) and deep learning molecular dynamics (DLMD) simulations, we demonstrate that polar domains can be created and manipulated in twisted bilayers of ferroelectric $CuInP_2S_6$, as a result of interfacial ferroelectric (antiferroelectric) coupling in AA (AB) stacked region. Unlike the topological polar vortex and skyrmions observed in superlattices of $(PbTiO_3)_n/(SrTiO_3)_n$ and sliding bilayers of BN and $MoS_2$, the underlying mechanism of polar domain formation in this system arises from stacking-dependent energy barriers for ferroelectric switching and variations in switching speeds under thermal perturbations. Notably, the thermal stability and polarization lifetimes are highly sensitive to twist angles and temperature, and can be further manipulated by external electric fields and strain. Through multi-scale simulations, our study provides a novel approach to exploring how twist angles influence domain evolution and underscores the potential for controlling local polarization in ferroelectric materials via rotational manipulation.

# 1 Introduction

Polar domains or vortices, referring to regions with uniformly aligned electric dipoles, represent a vibrant research frontier in condensed matter physics, as evident from the discoveries of domain-induced negative capacitance, [1-4] topological skyrmionic states, [5-8] and unconventional phase transitions. [9-12] Understanding the microscopic and macroscopic characteristics of these topological phenomena, precisely manipulating these domains, as well as revealing critical factors influencing their stability and morphology, are of fundamental and practical interest. [13] Traditionally, stress-strain effects have been employed to facilitate the formation of polar topologies and the annihilation of polarization domain. For example, lattice mismatch strain in heterostructures $(PbTiO_3)_n/(SrTiO_3)_n$ has enabled the realization of stable polarization vortices or polar skyrmions within ferroelectric bubble domains at room temperature. [5, 14] This behaviour is driven by the competition between electrostatic and strain boundary conditions, which inhibits the formation of a homogeneous ferroelectric ground state. The strain engineering techniques are recently also utilized to induce localized spontaneous polarization switching in van der Waals (vdW) ferroelectric materials, such as $In_2Se_3$ and $CuInP_2S_6$ (CIPS). Localized bending and bubbling under high strain conditions lead to the formation of stable polarization domains. [15-17] However, it is noticed that strain engineering has its inherent limitations. The ability to precisely tune strain gradients via mechanical boundary conditions is often restricted, complicating control over polarization switching. [18, 19] Furthermore, in heterostructures, imperfections at the surfaces and interfaces significantly affect interface coupling, [20] with defects and dislocations impeding domain nucleation or influencing domain walls' motion, thereby reducing overall domain stability. [21, 22]

Associated with the discovery of superconductivity in twisted bilayer graphene, twist deformation has recently emerged as an effective approach to regulate interlayer coupling in low-dimensional ferroelectric materials. [23-25] The induced Moiré superlattices can enhance interlayer quantum coupling and electron interactions to form stable polar topologies. For example, tunable topological vortices have been observed in twisted bilayer $MoS_2$ structures, [26, 27] while rotational polarization textures are generated in twisted freestanding $BaTiO_3$ layers due to nanoscale strain modulation. [19] Additionally, alternating polar domains with anti-aligned dipoles have been identified in twisted $WSe_2$. [28, 29] These discoveries highlight the potential of twist-induced Moiré superlattices for the electrical control of ferroelectric domains. Nevertheless, experimental challenges persist, particularly in achieving high-stable ferroelectric domain and understanding the atomic-level coupling between twist angles and polarization switching. [30-32]

Here, we predict the existence of stable polar domains in bilayer twisted CIPS and investigate their domain evolution dynamics under external driving forces through joint density functional theory (DFT) and molecular dynamics (MD) simulations. Static DFT calculations reveal that the stacking order in bilayer CIPS modifies the interlayer electron distribution, resulting in a lower polarization switching barrier for the ferroelectric (FE) to antiferroelectric (AFE) transition in the AB stacking compared with AA stacking. As a result, FE/AFE domains are localized around AA/AB stacking regions in twisted CIPS bilayers, as seen from large-scale MD simulations based on the developed deep learning potential. [33-35] The thermal stability of the ferroelectric domain and polarization lifetimes are highly sensitive to twist angles and temperature. They can be further enhanced or weakened through external electric fields and strain. These results not only deepen the understanding of ferroelectric domain formation and manipulation in low-dimensional systems but also highlight the potential of 2D ferroelectric materials in next-generation electronic devices, particularly in data storage and sensor technologies.

# 2 Results and Discussion

## *2.1 Switching Barrier Influenced by Stacking*

2D CIPS is a vdW ferroelectric material with experimentally confirmed room-temperature stability. [36] The spontaneous polarization originates from the broken structural symmetry along the out-of-plane direction, it is in ferroelectric phase when Cu ions are located on the top surface. The polarization reversal can be achieved with the vertical displacement of the Cu ions within the octahedron. [37] The large energy barrier (~200 meV) associated with Cu migration ensures the stability of the ferroelectric state at room temperature. [38] When one CIPS layer is stacked atop another, the resulting bilayer can exhibit either a ferroelectric or antiferroelectric phase, depending on the relative positioning of the Cu cations in the two layers. In the antiferroelectric phase, the Cu cations in the upper and lower layers are aligned in a head-to-head configuration (Figure 1a), with the polarizations of the two layers pointing in opposite directions. In contrast, the Cu cations adopt a head-to-foot configuration in the ferroelectric phase, resulting in an enhanced overall polarization.

A phase transition between ferroelectric and antiferroelectric states can be achieved when the Cu cation in one layer moves from the top to the bottom site within the octahedron. Within a 2×1 supercell, the polarization reversal exhibits a bimodal behaviour with two energy barriers due to the presence of metastable structure c/c' (Figure 1a). More interesting, the energy barriers are stacking-dependent, as shown in Figure 1b. Under AA-stacking configuration (each atom in the upper layer is directly aligned with the corresponding atom in the lower layer), the energy barriers for a Cu ion motion within the upper CIPS layer are 200 meV (a → c transition) and 231 meV (c → e transition) respectively. As a comparison, the values in the AB-stacked configuration (Cu/In atoms in the upper layer are located directly above the S atoms in the lower layer) are 212 meV and 208 meV, respectively. This implies that, although the initiation of Cu migration for polarization reversal in AB stacking is relatively more difficult comparing with AA configuration (200 vs 212 meV), the following displacement and associated polarization reversal will be much easier (231 vs 208 meV). The difference originates from the stacking-dependent interfacial interactions and electron distributions. DFT calculations with vdW corrections indicate that the interlayer distance for AA/AB stacking is 3.24/3.18 Å. As a result, comparing with AA stacking, more holes in AB structure are accumulated at the interlayer interface and more electrons gather between the layers (Figure 1b). This net charge buildup creates an internal electric field between the layers, which facilitates the movement of Cu cations, leading to an overall lower polarization switching barrier for AB stacking. The stacking dependent switching barrier in bilayer CIPS implies the potential to induce ferroelectric domain and coexisted FE/AFE phase when it undergoes interfacial twist to create the Morié pattern.

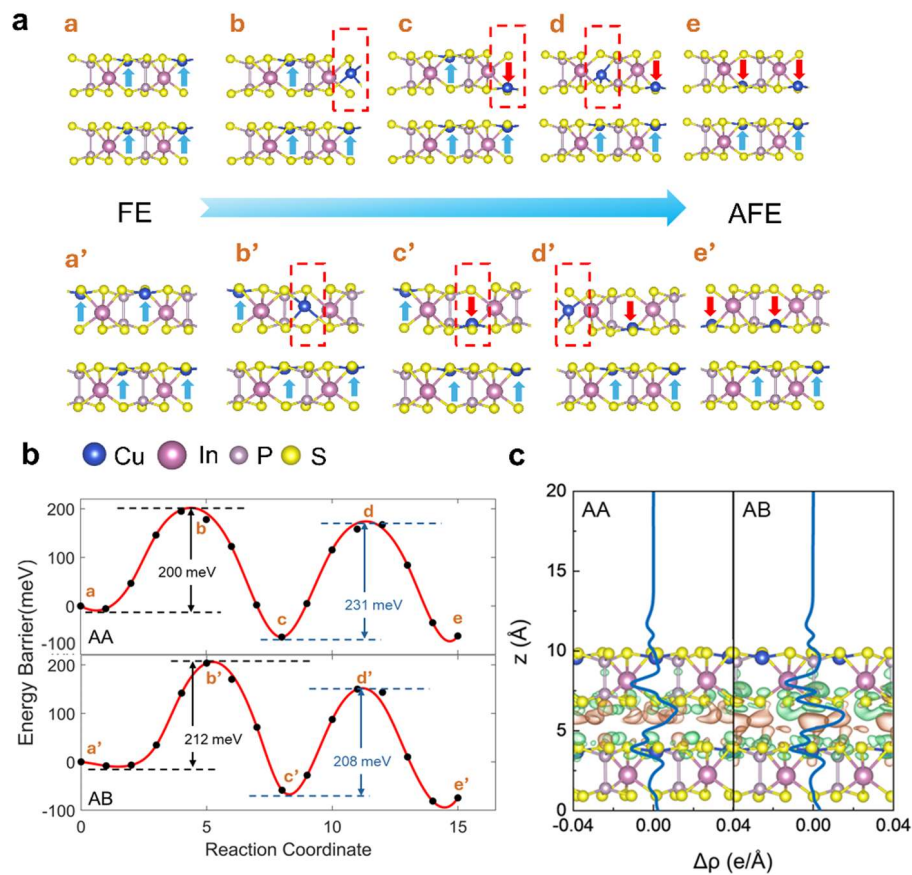

**Figure 1. DFT results for different stacking orders in bilayer CIPS.** a) The minimum energy paths for Cu displacement in AA & AB stacked 2×1 bilayers. b) The energy barriers for FE ↔AFE transitions in AA & AB stacked bilayers. c) Planar-averaged electron density difference Δρ(z) for CIPS as a function of the z coordinate which is perpendicular to the atomic planes. The green and orange areas indicate electron depletion and accumulation, respectively.

## *2.2 Deep-potential Predicted Switching Properties*

Static DFT simulations can accurately describe the energetics and electronic structure of the system, however their inherent limitations related to high computational cost prevent them from mimicking complex Moiré superlattice systems and studying thermal disturbance and dynamic process which more effectively reflect the real-world impacts. Here, the force fields of CIPS are developed using the DeepMD-kit [33] to enable the large-scale MD simulations, with a training dataset derived from the open-source data [39] as well as our DFT calculations. Energies, forces, and atomic coordinates for each high-symmetry stacking configuration are calculated using vdW-corrected DFT approach to capture the delicate interlayer interactions. These values are then used as inputs for the machine learning model, ensuring the accuracy of subsequent DeepMD simulations in reproducing the stacking-dependent interlayer interactions. The stacking order within local regions of the Moiré pattern was determined by laterally shifting the top layer, as illustrated in Supporting Information Figure S1. Our benchmark calculations indicate that the developed force field performs remarkably well in both energy and force predictions, with the mean absolute error (MAE) in energy of around 1.26 meV/atom (Supporting Information Figure S2).

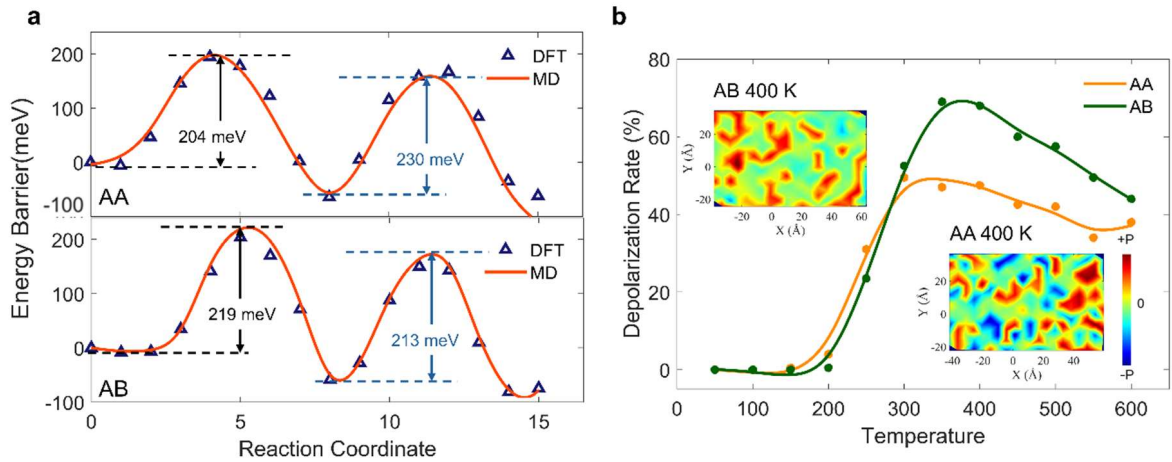

**Figure 2. DLMD results for different stacking orders in bilayer CIPS.** a) The switching barrier for FE ↔ AFE phase transition in AA and AB structures, as reproduced from DLMD simulations. The triangles represent the NEB results from the DFT calculations, while the line indicates the barrier calculation curve from the DLMD potential. b) Depolarization rate of AA and AB stacked structures as a function of temperature at 100 ps. The insets show the polarization distribution diagrams of both structures at 400 K.

The developed Deep Potential (DP) force field demonstrates exceptional accuracy in predicting both energies and forces (Supporting Information Figure S2 & S3). It can predict the energies of the metastable AA and ground-state AB stacking configurations with remarkable precision, yielding errors of only 0.006% and 0.02%, respectively, relative to DFT values (Supporting Information Figure S4). The transition process of Cu cation movement is also accurately reproduced by the DP force field, with energy profiles from both methods being nearly identical. Specifically, the switching barriers are 204 and 230 meV for the AA-stacked structure, 219 and 213 meV for the AB-stacked structure respectively. This prefect agreement highlights the high accuracy and reliability of the developed potential and its capability to model the dynamic evolution of phase transitions from the FE state to the AFE state under thermal disturbances (Figure 2a).

Given the lower energy barrier, ferroelectric polarization in the AB stacking configuration is expected to depolarize more easily under external thermal disturbance. To test this hypothesis, we conducted DeepMD simulations using 10 × 10 × 1 supercells, initializing the two layers with a head-to-foot polarized configuration to study the domain evolution dynamics in both stacking structures (Figure 2b). The results indicate that the polarizations of both configurations remain stable and unreversed until the temperature increases to 150 K, suggesting that thermal disturbances at this level are insufficient to overcome the energy barrier for ferroelectric switching. As the temperature rises further, the depolarization rate, defined as the average percentage of AFE domain formation (depolarization) per picosecond, diverges between the two configurations. Due to its lower resistance to thermal perturbations (the first energy barrier is relatively smaller), AA-stacked structure exhibits a slightly faster depolarization rate than AB stacking within the temperature range of 150 – 270 K, facilitated by the increased likelihood of intermediate-state formation. Under higher temperature (above 270 K), the AB-stacked structure, benefiting from a lower secondary energy barrier, undergoes a significant increase in depolarization rate, leading to the formation of larger AFE domains, as shown in the polarization distribution at 400 K and 100 ps. In contrast, the AA-stacked structure exhibits limited AFE domain expansion, with extensive polarization retention and reversal events. These results validate the static and dynamic stacking-dependent polarization switching behaviours observed in both DFT and DeepMD simulations, further implying the potential to create and manipulate polarization domains in complex systems, such as twisted bilayer structures.

## *2.3 Twist-induced Polarization Domain and Thermal Stability*

Here, a twisted CIPS bilayer structure is built to explore the influence of stacking on polarization domain formation, manipulation, and the thermal stability of these domains. Four models with twist angles of 2.28°, 3.48°, 4.41°, and 5.08° (see Figure S5) were constructed for large-scale DeepMD simulations. The resulting Moiré superlattice exhibits a coexistence of AA and AB stackings, with a distinct boundary separating the regions. Based on recent experimental findings (the interlayer is in ferroelectric coupling in bulk or multiple layer CIPS [36]) and stacking-dependent stability as above, AA stacking region will be trapped in local potential wells and keep as FE coupling while AB will be more likely to be switched into AFE coupling as temperature increase. Thus, the stacking-dependent interlayer coupling induces polarization domains in the AA-stacked region, but non-polarized (antiferroelectric coupling) in the AB-stacked region (Figure S6). The ferroelectric domains in the AA region depend on the twist angle, where a decrease in the twist angle leads to an increase in the antiferroelectric domain size. The average radius of non-polarized domains decreases from 53 Å to 22 Å as the twist angle increases from 2.28° to 5.08° (Figure 3a).

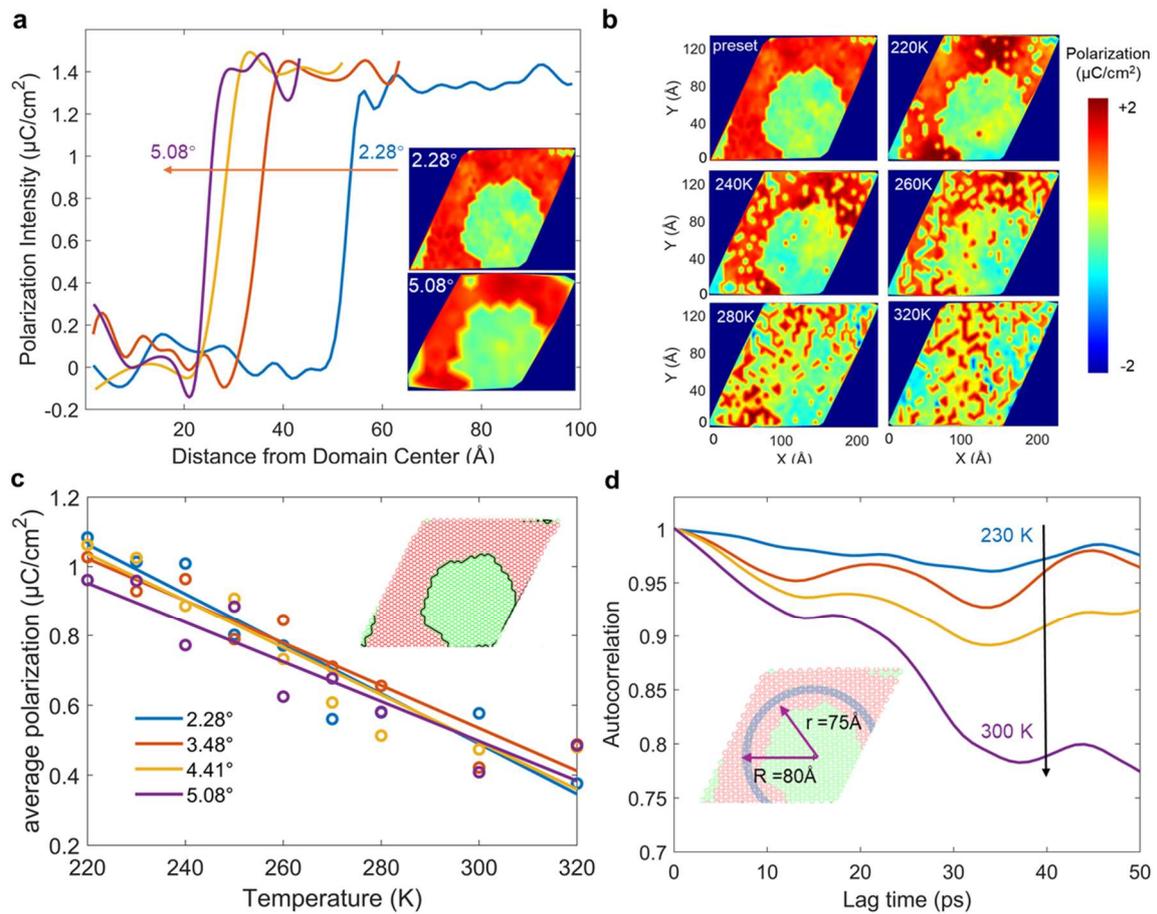

**Figure 3. Ferroelectric domain and dynamics in moiré twist.** a) Polarization distribution for twisted bilayers under static conditions. b) Heatmap of the polarization distribution at various temperatures for a twisted bilayer with a twist angle of 2.28°. c) Average polarization as a function of temperature. The illustration highlights the boundary of the polarization domain at 0 K, with the red region representing the initial ferroelectric (FE) domain. d) Temperature-dependent correlation function of polarization, with the polarization averaged over the blue region.

DeepMD simulations were conducted on twisted models at external temperatures to investigate the thermal stability of polar domains. Snapshots of polarization domain evolution at 100 ps for the 2.28° twisted bilayer under various temperatures are shown in Figure 3b and Figure S7-10. At 220 K, while thermal perturbations increase the atomic energy, it remains insufficient to overcome the energy barrier required for polarization reversal in the AA- and AB-stacked configurations. As a result, both FE and AFE domains maintain relatively stable. At 260 K, however, polarized domains undergo significant disruption, part of AA configurations are transited from the FE to AFE state, as thermal fluctuations destabilize the FE domain. As temperature further increases, both AFE and FE domains is destroyed in the twisted structure, no clear domain boundary can be observed (at 320 K). The average polarization as a function of temperature was calculated across different twist angles (Figure 3c) to quantify impacts of thermal disturbance on the domain stability. It decreases with increasing temperature in a nearly linear trend, indicating a uniform thermodynamic response of AA-stacked structures with the same twist angle to thermal fluctuations below 320 K. Notably, polarization regions associated with smaller twist angles exhibit higher thermodynamic stability, whereas configurations with larger twist angles are more prone to disruptions from thermal fluctuations (Figure 3c).

Its lifetime is examined by the autocorrelation of the polarization distribution function within the domain relative to the initial states. As shown in Figure 3d, the autocorrelation remains nearly 100% at 230 K after 50 ps, indicating the polarization state keeping unchanged for 50 ps, namely its high stability over time. However, obvious polarization reversal can be observed within the domain at 300 K, as evidenced by a significant decrease in autocorrelation, approximately 77% dropping at 50 ps. This reduction in polarization lifetime at higher temperatures reflects the difficulty in maintaining stable domain walls and nuclei, as seen in the disrupted FE and AFE domains at 320 K (Figure 3b). As such, our findings demonstrate that both temperature and twist angle play a crucial role in determining the stability of polarized domains.

## *2.4 Modulation of Polarization Domain by External Fields*

External electric field is commonly used to manipulate ferroelectric domains, as demonstrated in twisted $MoS_2$, $WSe_2$ and BN. [27, 29, 40] Here the manipulation of electric field on polarization domain, focusing on stability, lifetime, and size is studied in twisted CIPS bilayers. DFT simulations (Supporting Information Figure S11) revealed that the first energy barriers for FE-AFE phase transitions in both AA- and AB-stacked bilayers linearly increase or decrease under positive or negative electric fields, respectively. This behaviour results from the fact that a positive/negative electric field suppresses/facilitates Cu ion migration due to Coulombic interactions. It also implies the effective manipulation of electric field on ferroelectric domain in twisted bilayer. Here, polarization domain evolution in twisted bilayers (2.28° and 4.41°) was studied from DeepMD simulations at 220 K under vertical electric fields ranging from ±0.1 to ±0.9 V/Å. The effect of the electric field was modelled as an equivalent force exerted on each atom, based on Born effective charge (Section 3 in Supporting Information). Variations in Born effective charges among ions, along with induced interatomic interactions, enable the modulation of the polarization switching barrier under external electric fields. Consistent with DFT findings, the energy barrier for polarization reversal decreases under a negative electric field, leading to depolarization and destabilization of the FE domain. For instance, partial depolarization within the FE domain transitions to AFE coupling at E = −0.1 V/Å. As the field strength increases, the proportion and area of AFE-coupled domains expand, resulting in larger domains and domain fusion. At E = −0.8 V/Å, a significantly enlarged central AFE domain emerges, disrupting the boundary between FE and AFE domains, effectively dividing the FE domain (Figure 4a, Supporting Information Figure S12-13). Conversely, under a positive electric field aligned with the polarization direction in the AA region, the FE domain remains stable. The Cu cations in the AFE domain shift toward the polarization direction, forming the polarized islands. Due to the higher potential barrier for AFE-to-FE transition compared to the depolarization process (Figure 1a), the AFE domain exhibits better stability under the same field intensity. Even at E = 0.8 V/Å, the AFE domain remains relatively undisturbed. To quantitatively describe these findings, Figure 4a illustrates the percentage of structures in which polarization within the FE domain remains unchanged under an external electric field over 100 ps. The results show a nearly linear decrease (increase) in ferroelectric domain size in the AA region under negative (positive) electric fields, respectively.

The electric field not only influences domain size but also significantly impacts its lifetime, a key indicator of ferroelectric domain's stability. To quantify this effect, we calculated the autocorrelation of polarization in the AA-stacked region, where the interlayer coupling is ferroelectric. As shown in Figure 4b, when the electric field intensity ranges from −0.1 to −0.5 V/Å, the autocorrelation function remains nearly constant at 1, suggesting that the polarization state is stable over time. However, at −0.8 V/Å, the autocorrelation function decreases significantly, reflecting transient polarization within the FE domain and the disruption of stable domain structures. In contrast, under a positive electric field, the autocorrelation function keeps almost unchanged up to 0.8 V/Å, indicating the enhanced stability of the FE domain in the AA-stacked region (see Figure S12-13).

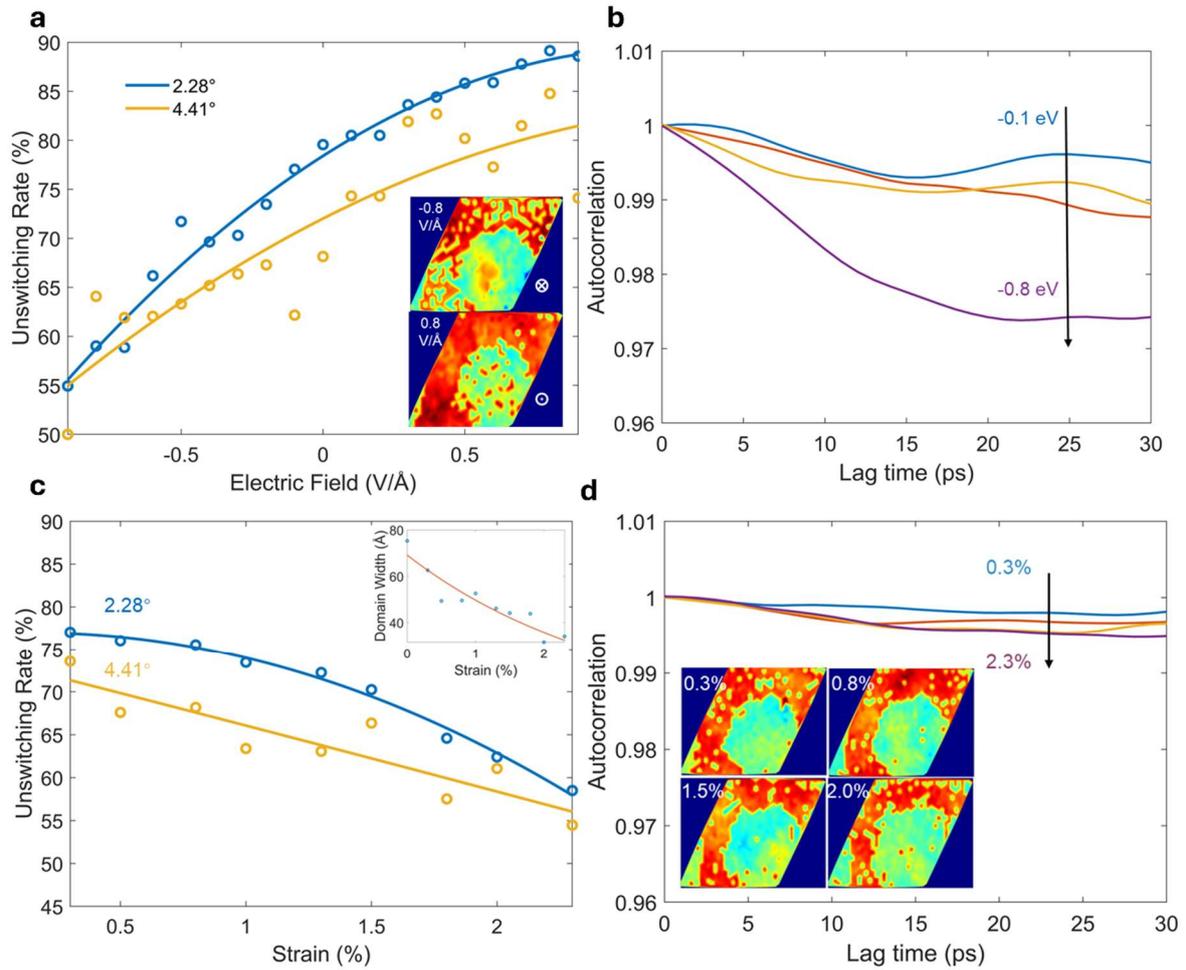

**Figure 4. Polarization domain behaviour in moiré twist under external field.** a) Percentage of structures retaining polarization unchanged in a twisted Moiré bilayer CIPS as a function of a perpendicular electric field at 220 K. Insets show heatmaps of the polarization distribution under electric fields of ±0.8 V/Å for a structure with a twist angle of 2.28°. b) Correlation function of polarization as a function of electric field strength. c) Percentage of structures retaining polarization in a twisted Moiré bilayer CIPS as a function of strain at 220 K. Inset depicts changes in ferroelectric (FE) domain width under strain for the 2.28° model. d) Strain-dependent correlation function of polarization. The illustrations include heatmaps of the polarization distribution at strain levels of 0.3%, 0.8%, 1.5%, and 2.0%.

In addition to electric fields, tensile strain along the <100> direction can also influence the polarization domain in the twisted structures. It is observed that the width of the FE domain is shrinking notably at 220 K (Figures. 4c, 4d, Supporting Information Figure S14-15), while the domain stability is weakened by external strain as evidenced by the decreasing percentage of the structures that can retain polarization. The impact of strain is twisting angle dependent, the structures with larger twist angles exhibit lower resilience against to strain, a more rapid decay of polarized domains is found (Figure 4c). The polarization lifetime, however, is less pronounced than that of thermal fluctuations or electric fields. Although the autocorrelation decreases slightly with increasing strain, it remains close to 1, indicating relatively stable polarization over time (Figure 4d). This is because tensile strain induces only minor atomic displacements, unlike the significant atomic rearrangements caused by thermal fluctuations or electric fields. Consequently, the overall polarization within the FE domain is largely stabilized, despite a visible reduction in domain width.

## 3 Conclusion

In conclusion, a joint DFT and DeepMD simulations were employed to investigate polarization domain evolution in twisted CIPS bilayers. Our DFT calculations reveal that bilayer CIPS with different stacking configurations exhibit distinct polarization switching barriers due to stacking-dependent interlayer interactions and electron transfer, providing the theoretical foundation for our findings. Using machine learning methods and the developed DP potential, large-scale molecular dynamics simulations confirm the presence of ferroelectric domains in twisted CIPS bilayers, where the AA and AB regions of the moiré structure exhibit interlayer FE and AFE coupling, respectively. The polarization domains and their stabilities are shown to be significantly influenced by thermal disturbances and externally applied electric fields, while demonstrating strong resistance to tensile strain. Our research offers a novel pathway for creating and manipulating polarization domains.

# 4 Experimental Section

**DFT Calculation for DL Potential Database and Switching Barrier**

All DFT calculations were performed using the Vienna Ab Initio Simulation Package (VASP).[41, 42] The exchange-correlation functional was treated with the optB86 approximation,[43] incorporating van der Waals corrections. The interaction between valence electrons and ion cores was described by the projected augmented wave (PAW) method and the Brillouin zone was sampled using the Monkhorst−Pack mesh with a reciprocal space resolution of 0.20 Å$^{-1}$. The energy cutoff for the plane wave basis was set to 500 eV. All atoms were allowed to relax until the Hellmann-Feynman forces were smaller than 0.001 eV/Å, and the convergence criterion for the electronic self-consistent loop was set to 10$^{-6}$ eV. A vacuum region of 20 Å was included to eliminate the artifactual interactions between periodic images. To replicate the domain-wall motion, the climbing image nudged elastic band (CI-NEB) method was employed to determine the minimum energy path from the FE state to the AFE state.[44, 45]

**DL Potential Training and MD Simulation**

The DeepMD-kit [33] was employed as the training package with the descriptor type set to "$se\_e2\_a$". A cutoff value of 6.0 Å and a smoothed radius of 2.0 Å were chosen to ensure descriptor smoothness. The embedding neural network used for descriptor training had a size of (25,50,100), while the fitting neural network had 240 neurons in each of its three layers. In loss function setting, the weight $P_e$ was progressively increased from 0.02 to 2, while the weight $P_f$ was gradually reduced from 1000 to 1. Additionally, the virial weight $P_v$ was 0. The learning rate was initially set at 0.001 and decayed to 3.51e-08 after 10000 steps. The entire training process spanned 5,000,000 steps. The loss function [35] in Deepmd-kit, including the physic information of force and virial. Shows in the following equation:

$$L\left(p_e, p_f, p_v\right) = \frac{p_e}{N}\Delta E^2 + \frac{p_f}{3N}\sum_i |\Delta F_i|^2 + \frac{p_v}{9N}||\Delta v||^2 \qquad (1)$$

Where the three items $\Delta E$, $\Delta F$ and $\Delta v$ mean root mean square (RMS) error in energy, force, and virial. The three parameters $p_e, p_f$ and $p_v$ is the weights of this three information.

The MD simulations were performed using the Large-scale Atomic/Molecular Massively Parallel Simulator (LAMMPS). [46] Periodic boundary conditions were applied in the in-plane directions, while the out-of-plane direction was set to be aperiodic. The simulations were conducted in the canonical ensemble (NVT), with temperature control maintained via the Nosé-Hoover thermostat. [47, 48]

**AUTHOR CONTRIBUTIONS**

L.K. conceived the idea and supervised the research with Y.G.. D.B. designed the MD simulation process and models. D.B. trained the DLMD potential and conducted the MD simulations, assisted by Y.N.. J.L. executed the DFT simulations. D.B. analyzed the data with contributions from J.L., L.K., A.K., and D.Q.. D.B. composed the manuscript with input from J.L. and L.K.. All authors have given approval to the final version of the manuscript. D.B. and J.L. have made equal contributions to this work and can be recognized as co-first authors.

**ACKNOWLEDGMENTS**

The authors acknowledge the grants of high-performance computer (HPC) time from the computing facility at the Queensland University of Technology (QUT), as well as the computational resources from the Pawsey Supercomputing Centre and Australian National Computational Infrastructure (NCI) supported by the Australian Government. L.K. gratefully acknowledges financial support by the ARC Discovery Project (DP230101904 and DP240103085).

# Reference

[1] A.K. Yadav, K.X. Nguyen, Z. Hong, P. García-Fernández, P. Aguado-Puente, C.T. Nelson, S. Das, B. Prasad, D. Kwon, S. Cheema, Nature 2019, 565, 468-471.
[2] H.W. Park, J. Roh, Y.B. Lee, C.S. Hwang, Adv. Mater. 2019, 31, 1805266.
[3] L. Tu, X. Wang, J. Wang, X. Meng, J. Chu, Advanced Electronic Materials 2018, 4, 1800231.
[4] C. Zhou, Y. Chai, Advanced Electronic Materials 2017, 3, 1600400.
[5] S. Das, Y. Tang, Z. Hong, M. Gonçalves, M. McCarter, C. Klewe, K. Nguyen, F. Gómez-Ortiz, P. Shafer, E. Arenholz, Nature 2019, 568, 368-372.
[6] L. Han, C. Addiego, S. Prokhorenko, M. Wang, H. Fu, Y. Nahas, X. Yan, S. Cai, T. Wei, Y. Fang, Nature 2022, 603, 63-67.
[7] L. Gao, S. Prokhorenko, Y. Nahas, L. Bellaiche, Phys. Rev. Lett. 2024, 132, 026902.
[8] X. Guo, L. Zhou, B. Roul, Y. Wu, Y. Huang, S. Das, Z. Hong, Small Methods 2022, 6, 2200486.
[9] I.I. Naumov, L. Bellaiche, H. Fu, Nature 2004, 432, 737-740.
[10] S. Das, Z. Hong, V. Stoica, M. Gonçalves, Y.-T. Shao, E. Parsonnet, E.J. Marksz, S. Saremi, M. McCarter, A. Reynoso, Nature materials 2021, 20, 194-201.
[11] Z. Hong, L.-Q. Chen, Acta Materialia 2018, 152, 155-161.
[12] Z. Hong, L.-Q. Chen, Acta Materialia 2019, 164, 493-498.
[13] J.F. Scott, C.A. Paz de Araujo, Science 1989, 246, 1400-1405.
[14] Q. Zhang, L. Xie, G. Liu, S. Prokhorenko, Y. Nahas, X. Pan, L. Bellaiche, A. Gruverman, N. Valanoor, Adv. Mater. 2017, 29, 1702375.
[15] D. Bai, Y. Nie, J. Shang, J. Liu, M. Liu, Y. Yang, H. Zhan, L. Kou, Y. Gu, Nano Lett. 2023, 23, 10922-10929.
[16] C. Chen, H. Liu, Q. Lai, X. Mao, J. Fu, Z. Fu, H. Zeng, Nano Lett 2022, 22, 3275-3282.
[17] Y. Yang, H. Zong, J. Sun, X. Ding, Adv Mater 2021, 33, e2103469.
[18] G. Catalan, A. Lubk, A. Vlooswijk, E. Snoeck, C. Magen, A. Janssens, G. Rispens, G. Rijnders, D.H. Blank, B. Noheda, Nature materials 2011, 10, 963-967.
[19] G. Sánchez-Santolino, V. Rouco, S. Puebla, H. Aramberri, V. Zamora, M. Cabero, F. Cuellar, C. Munuera, F. Mompean, M. Garcia-Hernandez, Nature 2024, 626, 529-534.
[20] X. Liu, M.C. Hersam, Adv. Mater. 2018, 30, 1801586.
[21] R. Luo, M. Gao, C. Wang, J. Zhu, R. Guzman, W. Zhou, Adv. Funct. Mater. 2024, 34, 2307625.
[22] R. He, B. Zhang, H. Wang, L. Li, T. Ping, G. Bauer, Z. Zhong, arXiv preprint arXiv:2212.14203 2022.
[23] Y. Cao, V. Fatemi, A. Demir, S. Fang, S.L. Tomarken, J.Y. Luo, J.D. Sanchez-Yamagishi, K. Watanabe, T. Taniguchi, E. Kaxiras, Nature 2018, 556, 80-84.
[24] K. Liu, L. Zhang, T. Cao, C. Jin, D. Qiu, Q. Zhou, A. Zettl, P. Yang, S.G. Louie, F. Wang, Nat. Commun. 2014, 5, 4966.

[25] M. Van Wijk, A. Schuring, M. Katsnelson, A. Fasolino, Phys. Rev. Lett. 2014, 113, 135504.
[26] C.S. Tsang, X. Zheng, T. Yang, Z. Yan, W. Han, L.W. Wong, H. Liu, S. Gao, K.H. Leung, C.-S. Lee, Science 2024, 386, 198-205.
[27] A. Weston, E.G. Castanon, V. Enaldiev, F. Ferreira, S. Bhattacharjee, S. Xu, H. Corte-León, Z. Wu, N. Clark, A. Summerfield, Nature nanotechnology 2022, 17, 390-395.
[28] K. Ko, A. Yuk, R. Engelke, S. Carr, J. Kim, D. Park, H. Heo, H.-M. Kim, S.-G. Kim, H. Kim, Nature Materials 2023, 22, 992-998.
[29] M. Van Winkle, N. Dowlatshahi, N. Khaloo, M. Iyer, I.M. Craig, R. Dhall, T. Taniguchi, K. Watanabe, D.K. Bediako, Nature Nanotechnology 2024, 1-7.
[30] K. Kim, M. Yankowitz, B. Fallahazad, S. Kang, H.C. Movva, S. Huang, S. Larentis, C.M. Corbet, T. Taniguchi, K. Watanabe, Nano Lett. 2016, 16, 1989-1995.
[31] S. Grover, M. Bocarsly, A. Uri, P. Stepanov, G. Di Battista, I. Roy, J. Xiao, A.Y. Meltzer, Y. Myasoedov, K. Pareek, Nature physics 2022, 18, 885-892.
[32] A. Uri, S. Grover, Y. Cao, J.A. Crosse, K. Bagani, D. Rodan-Legrain, Y. Myasoedov, K. Watanabe, T. Taniguchi, P. Moon, Nature 2020, 581, 47-52.
[33] H. Wang, L. Zhang, J. Han, W. E, Comput Phys Commun 2018, 228, 178-184.
[34] Y. Zhang, H. Wang, W. Chen, J. Zeng, L. Zhang, H. Wang, E. Weinan, Comput Phys Commun 2020, 253, 107206.
[35] L. Zhang, J. Han, H. Wang, R. Car, W. E, Phys. Rev. Lett. 2018, 120, 143001.
[36] F. Liu, L. You, K.L. Seyler, X. Li, P. Yu, J. Lin, X. Wang, J. Zhou, H. Wang, H. He, Nat. Commun. 2016, 7, 1-6.
[37] D. Seleznev, S. Singh, J. Bonini, K.M. Rabe, D. Vanderbilt, Phys. Rev. B 2023, 108, L180101.
[38] H. Liu, S. Yu, Y. Wang, B. Huang, Y. Dai, W. Wei, The Journal of Physical Chemistry Letters 2022, 13, 1972-1978.
[39] R. He, H. Wang, F. Liu, S. Liu, H. Liu, Z. Zhong, Phys. Rev. B 2023, 108, 024305.
[40] K. Yasuda, X. Wang, K. Watanabe, T. Taniguchi, P. Jarillo-Herrero, Science 2021, 372, 1458-1462.
[41] G. Kresse, J. Hafner, Phys. Rev. B 1993, 48, 13115.
[42] G. Kresse, J. Furthmüller, Phys. Rev. B 1996, 54, 11169.
[43] J. Klimeš, D.R. Bowler, A. Michaelides, Physical Review B—Condensed Matter and Materials Physics 2011, 83, 195131.
[44] G. Henkelman, H. Jónsson, J. Chem. Phys. 2000, 113, 9978-9985.
[45] G. Henkelman, B.P. Uberuaga, H. Jónsson, J. Chem. Phys. 2000, 113, 9901-9904.
[46] S. Plimpton, J. Comput. Phys. 1995, 117, 1-19.
[47] S. Nosé, J. Chem. Phys. 1984, 81, 511-519.
[48] W.G. Hoover, Physical review A 1985, 31, 1695.